\documentclass[american,aps, prl, showpacs, reprint]{revtex4-1}
\usepackage[T1]{fontenc}
\usepackage[latin9]{inputenc}
\usepackage{babel}

\usepackage{amsthm}
\usepackage{amsmath}
\usepackage{graphicx}
\usepackage[unicode=true, 
 bookmarks=true,bookmarksnumbered=false,bookmarksopen=false,
 breaklinks=false,pdfborder={0 0 0},backref=false,colorlinks=false]
 {hyperref}
\hypersetup{
 pdfauthor={Alexander Streltsov, Hermann Kampermann, Dagmar Bruß}}
 
\makeatletter
\@ifundefined{textcolor}{}
{%
 \definecolor{BLACK}{gray}{0}
 \definecolor{WHITE}{gray}{1}
 \definecolor{RED}{rgb}{1,0,0}
 \definecolor{GREEN}{rgb}{0,1,0}
 \definecolor{BLUE}{rgb}{0,0,1}
 \definecolor{CYAN}{cmyk}{1,0,0,0}
 \definecolor{MAGENTA}{cmyk}{0,1,0,0}
 \definecolor{YELLOW}{cmyk}{0,0,1,0}
 }
\theoremstyle{plain}
\newtheorem{thm}{Theorem}

\usepackage{lmodern}
\usepackage{braket}

\makeatother

\begin{document}

\title{Linking quantum discord to entanglement in a measurement}

\author{Alexander Streltsov}

\email{streltsov@thphy.uni-duesseldorf.de}

\author{Hermann Kampermann}

\author{Dagmar Bruß}

\affiliation{Heinrich-Heine-Universit\"{a}t D\"{u}sseldorf, Institut f\"{u}r
Theoretische Physik III, D-40225 D\"{u}sseldorf, Germany}
\begin{abstract}
We show that a von Neumann measurement on a part of a composite quantum
system unavoidably creates distillable entanglement between the measurement
apparatus and the system if the state has nonzero quantum discord.
The minimal distillable entanglement is equal to the one-way information
deficit. The quantum discord is shown to be equal to the minimal partial
distillable entanglement that is the part of entanglement which is
lost, when we ignore the subsystem which is not measured. We then
show that any entanglement measure corresponds to some measure of
quantum correlations. This powerful correspondence also yields necessary
properties for quantum correlations. We generalize the results to
multipartite measurements on a part of the system and on the total
system.
\end{abstract}


\maketitle
Quantum entanglement is by far the most famous and best studied kind
of quantum correlation \cite{Horodecki2009}. One reason for this
situation is the fact that entanglement plays an important role in
quantum computation \cite{Nielsen2000}. It was even believed that
entanglement is the reason why a quantum computer can perform efficiently
on some problems which cannot be solved efficiently on a classical
computer. The situation started to change after a computational model
was presented which is referred to as {}``the power of one qubit''
with the acronym DQC1 \cite{Knill1998,Laflamme2002}. Here, using
a mixed separable state allows for efficient computation of the trace
of any $n$-qubit unitary matrix. This problem is believed to be not
solvable efficiently on a classical computer \cite{Laflamme2002,Datta2005}.
The fact that no entanglement is present in this model was one of
the main reasons why new types of quantum correlations were studied
during the past few years \cite{Ollivier2001,Henderson2001,Oppenheim2002,Daki'c2010}.
One of the measures of quantum correlations, the quantum discord \cite{Ollivier2001},
was considered to be the figure of merit for this model of quantum
computation \cite{Datta2008}.

In this Letter, we introduce an alternative approach to quantum correlations
via an interpretation of a measurement. In order to perform a von
Neumann measurement on a system $S$ in the quantum state $\rho^{S}$,
correlations between the system and the measurement apparatus $M$
must be created. As a simple example we consider a von Neumann measurement
in the eigenbasis $\left\{ \ket{i^{S}}\right\} $ of the mixed state
$\rho^{S}=\sum_{i}p_{i}\ket{i^{S}}\bra{i^{S}}$ with the eigenvalues
$p_{i}$. Correlations between the measurement apparatus $M$ and
the system are found in the final state of the total system $\rho_{\mathrm{final}}=\sum_{i}p_{i}\ket{i^{M}}\bra{i^{M}}\otimes\ket{i^{S}}\bra{i^{S}}$,
where $\ket{i^{M}}$ are orthogonal states of the measurement apparatus
$M$. In this state $\rho_{\mathrm{final}}$ the correlations between
$M$ and the system $S$ are purely classical, and no entanglement
is created. The situation changes completely if we consider \emph{partial}
von Neumann measurements; that is, they are restricted to a part of
the system. In our main result in Theorem \ref{thm:1} we will show
that in this case creation of entanglement is usually unavoidable.
We use this result to show the close connection of our approach to
the one-way information deficit \cite{Oppenheim2002} before we extend
our ideas to the quantum discord \cite{Ollivier2001} in Theorem \ref{thm:2}
and following.

If we consider bipartite quantum states $\rho^{AB}$, and von Neumann
measurements on $A$ with a complete set of orthogonal rank one projectors
$\Pi_{i}^{A}=\ket{i^{A}}\bra{i^{A}}$, $\sum_{i}\Pi_{i}^{A}=\openone_{A}$,
then the quantum discord is defined as \cite{Ollivier2001} \begin{eqnarray}
\delta^{\rightarrow}\left(\rho^{AB}\right) & = & S\left(\rho^{A}\right)-S\left(\rho^{AB}\right)+\min_{\left\{ \Pi_{i}^{A}\right\} }\sum_{i}p_{i}S\left(\rho_{i}\right),\label{eq:discord}\end{eqnarray}
with $p_{i}=\mathrm{Tr}\left[\Pi_{i}^{A}\rho^{AB}\Pi_{i}^{A}\right]$
being the probability of the outcome $i$, and $\rho_{i}=\Pi_{i}^{A}\rho^{AB}\Pi_{i}^{A}/p_{i}$
being the corresponding state after the measurement. The quantum discord
is nonnegative and zero if and only if the state $\rho^{AB}$ has
the form $\rho^{AB}=\sum_{i}p_{i}\ket{i^{A}}\bra{i^{A}}\otimes\rho_{i}^{B}$
with orthogonal states $\ket{i^{A}}$. Recently an interpretation
of the quantum discord was found using a connection to extended state
merging \cite{Cavalcanti,Madhok}. Another interpretation was given
earlier in \cite{Zurek2003}.

A closely related quantity is the one-way information deficit \cite{Oppenheim2002,Horodecki2005}.
For a bipartite state $\rho^{AB}$ it is defined as the minimal increase
of entropy after a von Neumann measurement on $A$:\begin{equation}
\Delta^{\rightarrow}\left(\rho^{AB}\right)=\min_{\left\{ \Pi_{i}^{A}\right\} }S\left(\sum_{i}\Pi_{i}^{A}\rho^{AB}\Pi_{i}^{A}\right)-S\left(\rho^{AB}\right),\label{eq:deficit-1}\end{equation}
where the minimum is taken over $\left\{ \Pi_{i}^{A}\right\} $ as
defined above Eq. (\ref{eq:discord}). The one-way information deficit
is non-negative and zero only on states with zero quantum discord.
It can be interpreted as the amount of information in the state $\rho^{AB}$,
which cannot be localized via a classical communication channel from
$A$ to $B$ \cite{Horodecki2005}.

\begin{figure}
\noindent \begin{centering}
\includegraphics[width=0.9\columnwidth]{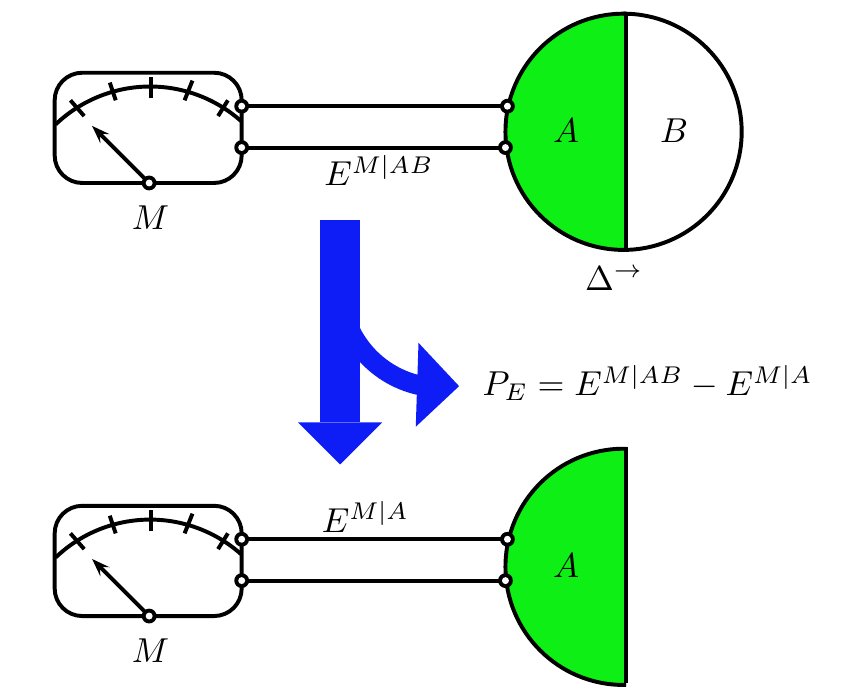}
\par\end{centering}

\caption{\label{fig:MAB}A measurement apparatus $M$ is used for a von Neumann
measurement on $A$ (green colored area), which is part of the total
quantum system $AB$. The measurement implies a unitary evolution
on the system $MA$, which can create entanglement $E^{M|AB}$ between
the apparatus and the system. The partial entanglement $P_{E}=E^{M|AB}-E^{M|A}$
quantifies the part of entanglement which is lost when ignoring $B$.}

\end{figure}

Given a bipartite quantum state $\rho^{AB}$, we recall that a partial
von Neumann measurement on $A$ can be described by coupling the system
in the state $\rho^{AB}$ to the measurement apparatus $M$ in a pure
initial state $\ket{0^{M}}$, $\rho_{1}=\ket{0^{M}}\bra{0^{M}}\otimes\rho^{AB}$,
and applying a unitary on the total state \cite{Schlosshauer2005},
$\rho_{2}=U\rho_{1}U^{\dagger}$. This situation is illustrated in
Fig. \ref{fig:MAB}. %
As we will consider only measurements on the subsystem $A$, the corresponding
unitary $U$ has the form $U=U_{MA}\otimes\openone_{B}$. In the following,
we will say that a unitary $U$ realizes a von Neumann measurement
$\left\{ \Pi_{i}^{A}\right\} $ on $A$, if for any quantum state
$\rho^{AB}$ holds: $\mathrm{Tr}_{M}\left[U\left(\ket{0^{M}}\bra{0^{M}}\otimes\rho^{AB}\right)U^{\dagger}\right]=\sum_{i}\Pi_{i}^{A}\rho^{AB}\Pi_{i}^{A}$.
The measurement outcome is then obtained by measuring the apparatus
$M$ in its eigenbasis. 

The entanglement between the apparatus $M$ and the system $AB$ in
the state $\rho_{2}$ will be called \emph{entanglement created in
the von Neumann measurement $\left\{ \Pi_{i}^{A}\right\} $} \emph{on}
$A$. Given a state $\rho^{AB}$, we want to quantify the minimal
entanglement created in a von Neumann measurement on $A$, minimized
over all complete sets of rank one projectors \emph{$\left\{ \Pi_{i}^{A}\right\} $}.
The minimal amount will be called $E_{\mathrm{meas}}$, and it will
depend on the entanglement measure used. In the following, the entanglement
measure of interest will be the distillable entanglement $E_{D}$,
which is defined in \cite{Bennett1996,Plenio2007}. Thus, we define
$E_{\mathrm{meas}}$ as follows: $E_{\mathrm{meas}}\left(\rho^{AB}\right)=\min_{U}E_{D}^{M|AB}\left(U\rho_{1}U^{\dagger}\right)$,
where the minimization is done over all unitaries which realize some
von Neumann measurement on $A$. Recalling the definition of the one-way
information deficit in (\ref{eq:deficit-1}), we present one of our
main results. 
\begin{thm}
\label{thm:1}If a bipartite state $\rho^{AB}$ has nonzero quantum
discord $\delta^{\rightarrow}\left(\rho^{AB}\right)>0$, any von Neumann
measurement on $A$ creates distillable entanglement between the measurement
apparatus and the total system $AB$. The minimal distillable entanglement
created in a von Neumann measurement on $A$ is equal to the one-way
information deficit: $E_{\mathrm{meas}}\left(\rho^{AB}\right)=\Delta^{\rightarrow}\left(\rho^{AB}\right)$.\end{thm}
\begin{proof}
As pointed out in \cite{Vedral2003}, the unitary $U$ must act on
states of the form $\ket{0^{M}}\otimes\ket{i^{A}}$ as follows: $U\left(\ket{0^{M}}\otimes\ket{i^{A}}\right)=\ket{i^{M}}\otimes\ket{i^{A}}$,
where $\left\{ \ket{i^{A}}\right\} $ is the measurement basis, and
$\ket{i^{M}}$ are orthogonal states of the measurement apparatus.
In general we can always write $\rho^{AB}=\sum_{i,j}\ket{i^{A}}\bra{j^{A}}\otimes O_{ij}^{B}$
with $O_{ij}^{B}$ being operators on the Hilbert space $\mathcal{H}_{B}$.
After the action of the unitary the state becomes $\rho_{2}=\sum_{i,j}\ket{i^{M}}\bra{j^{M}}\otimes\ket{i^{A}}\bra{j^{A}}\otimes O_{ij}^{B}$.
From \cite{Devetak2005} we know that the distillable entanglement
is bounded from below as $E_{D}^{M|AB}\left(\rho_{2}\right)\geq S\left(\rho_{2}^{AB}\right)-S\left(\rho_{2}\right)$
with $\rho_{2}^{AB}=\mathrm{Tr}_{M}\left[\rho_{2}\right]$, and the
von Neumann entropy $S\left(\rho\right)=-\mathrm{Tr}\left[\rho\log_{2}\rho\right]$.
We mention that the same inequality holds for the relative entropy
of entanglement defined in \cite{Vedral1997} as $E_{R}=\min_{\sigma\in S}S\left(\rho||\sigma\right)$
with the quantum relative entropy $S\left(\rho||\sigma\right)=-\mathrm{Tr}\left[\rho\log_{2}\sigma\right]+\mathrm{Tr}\left[\rho\log_{2}\rho\right]$;
see \cite{Plenio2000} for details. Noting that $\rho_{2}^{AB}=\sum_{i}\Pi_{i}^{A}\rho^{AB}\Pi_{i}^{A}$
and $S\left(\rho_{2}\right)=S\left(\rho_{1}\right)=S\left(\rho^{AB}\right)$
we see $E_{D}^{M|AB}\left(\rho_{2}\right)\geq S\left(\sum_{i}\Pi_{i}^{A}\rho^{AB}\Pi_{i}^{A}\right)-S\left(\rho^{AB}\right)$.
On the other hand, we know that $E_{R}$ is an upper bound on the
distillable entanglement \cite{Horodecki2000}. Consider the state
$\sigma=\sum_{i}\Pi_{i}^{M}\rho_{2}\Pi_{i}^{M}$, which is separable
with respect to the bipartition $M|AB$. From the definition of the
relative entropy of entanglement follows: $E_{R}^{M|AB}\left(\rho_{2}\right)\leq S\left(\rho_{2}||\sigma\right)$.
It can be seen by inspection that $S\left(\rho_{2}||\sigma\right)=S\left(\sum_{i}\Pi_{i}^{A}\rho^{AB}\Pi_{i}^{A}\right)-S\left(\rho^{AB}\right)$.
Thus we proved that $E_{D}^{M|AB}\left(\rho_{2}\right)=S\left(\sum_{i}\Pi_{i}^{A}\rho^{AB}\Pi_{i}^{A}\right)-S\left(\rho^{AB}\right)$
holds for any measurement basis $\left\{ \ket{i^{A}}\right\} $. If
we minimize this equation over all von Neumann measurements on $A$,
we get the desired result. 
\end{proof}
Note that from the above proof we conclude that $\underset{U}{\min}E_{D}^{M|AB}\left(U\rho_{1}U^{\dagger}\right)=\underset{U}{\min}E_{R}^{M|AB}\left(U\rho_{1}U^{\dagger}\right)$,
and thus there does not exist bound entanglement in a partial measurement.

The approach presented so far can also be applied to any other measure
of entanglement $E$, which satisfies the basic axiom to be nonincreasing
under local operations and classical communication (LOCC) \cite{Vedral1997}.
In this way we introduce the generalized one-way information deficit
as follows: \begin{eqnarray}
\Delta_{E}^{\rightarrow}\left(\rho^{AB}\right) & = & \min_{U}E^{M|AB}\left(U\rho_{1}U^{\dagger}\right),\label{eq:generalizedDeficit}\end{eqnarray}
where $U$ realizes a von Neumann measurement on $A$ and $\rho_{1}=\ket{0^{M}}\bra{0^{M}}\otimes\rho^{AB}$.
Using Theorem \ref{thm:1} it is easy to see that the generalized
one-way information deficit is zero if and only if the state $\rho^{AB}$
has zero quantum discord. This holds if $E$ is zero on separable
states only.

In the same way as different measures of entanglement capture different
aspects of entanglement, the correspondence (\ref{eq:generalizedDeficit})
can be used to capture different aspects of quantum correlations.
Let us demonstrate this by using the geometric measure of entanglement
$E_{G}$ \cite{Wei2003} on the right-hand side of (\ref{eq:generalizedDeficit}).
As the corresponding measure of quantum correlations, we obtain $\Delta_{E_{G}}^{\rightarrow}\left(\rho^{AB}\right)=\underset{\delta^{\rightarrow}\left(\sigma^{AB}\right)=0}{\min}\left\{ 1-F\left(\rho^{AB},\sigma^{AB}\right)\right\} $
with the fidelity $F\left(\rho,\sigma\right)=\left(\mathrm{Tr}\left[\sqrt{\sqrt{\rho}\sigma\sqrt{\rho}}\right]\right)^{2}$
\footnote{The proof will be given elsewhere%
}. The minimization is done over all states $\sigma^{AB}$ with zero
quantum discord. Thus, this measure captures the geometric aspect
of quantum correlations, similarly to the geometric measure of discord
presented in \cite{Daki'c2010}. 

The correspondence (\ref{eq:generalizedDeficit}) also implies that
certain properties of entanglement measures are transferred to corresponding
properties of quantum correlation measures. This will be demonstrated
in the following by finding a class of quantum operations which do
not increase $\Delta_{E}^{\rightarrow}$. This class cannot be equal
to the class of LOCC, since $\Delta_{E}^{\rightarrow}$ can increase
under local operations on $A$. This can be seen by considering the
classically correlated state $\rho_{cc}=\frac{1}{2}\ket{0^{A}}\bra{0^{A}}\otimes\ket{0^{B}}\bra{0^{B}}+\frac{1}{2}\ket{1^{A}}\bra{1^{A}}\otimes\ket{1^{B}}\bra{1^{B}}$
with $\Delta_{E}^{\rightarrow}\left(\rho_{cc}\right)=0$. Using only
local operations on $A$ it is possible to create states with nonzero
deficit $\Delta_{E}^{\rightarrow}$. Demanding that the subsystem
$A$ is unchanged, we are left with quantum operations on $B$ only.
In the following we will show that $\Delta_{E}^{\rightarrow}$ does
not increase under arbitrary quantum operations on $B$, denoted by
$\Lambda_{B}$: \begin{eqnarray}
\Delta_{E}^{\rightarrow}\left(\Lambda_{B}\left(\rho^{AB}\right)\right) & \leq & \Delta_{E}^{\rightarrow}\left(\rho^{AB}\right).\label{eq:deficitLO}\end{eqnarray}
Inequality (\ref{eq:deficitLO}) is seen to be true by noting that
the entanglement $E^{M|AB}$ does not increase under $\Lambda_{B}$,
as it does not increase under LOCC.

We can go one step further by noting that the distillable entanglement
is also nonincreasing on average under stochastic LOCC. This captures
the idea that two parties cannot share more entanglement on average,
if they perform local generalized measurements on their subsystems
and communicate the outcomes classically; see \cite{Plenio2007} for
more details. Defining the global Kraus operators describing some
LOCC protocol by $\left\{ V_{i}\right\} $ with $\sum_{i}V_{i}^{\dagger}V_{i}=\openone$,
the probability of the outcome $i$ is given by $q_{i}=\mathrm{Tr}\left[V_{i}\rho V_{i}^{\dagger}\right]$,
and the state after the measurement with the outcome $i$ is given
by $\sigma_{i}=V_{i}\rho V_{i}^{\dagger}/q_{i}$. Then for the distillable
entanglement \cite{Vidal2000} and the relative entropy of entanglement
holds \cite{Vedral1998} \begin{equation}
\sum_{i}q_{i}E\left(\sigma_{i}\right)\leq E\left(\rho\right).\label{eq:LOCC}\end{equation}
Inequality (\ref{eq:LOCC}) implies that the corresponding quantity
$\Delta_{E}^{\rightarrow}$ satisfies the related property \begin{equation}
\sum_{i}q_{i}\Delta_{E}^{\rightarrow}\left(\sigma_{i}^{AB}\right)\leq\Delta_{E}^{\rightarrow}\left(\rho^{AB}\right),\label{eq:deficitLO-1}\end{equation}
where $q_{i}$, $\sigma_{i}^{AB}$ are defined as above Eq. (\ref{eq:LOCC}),
and now $\left\{ V_{i}\right\} $ are Kraus operators describing a
local quantum operation on $B$. Inequality (\ref{eq:deficitLO-1})
is seen to be true by using (\ref{eq:LOCC}) in the definition (\ref{eq:generalizedDeficit}). 

In the following we will include the quantum discord $\delta^{\rightarrow}$
into our approach. We call the non-negative quantity \begin{equation}
P_{E}\left(\rho\right)=E^{M|AB}\left(\rho\right)-E^{M|A}\left(\rho^{MA}\right)\end{equation}
 the \emph{partial }e\emph{ntanglement}. It quantifies  the part of
entanglement which is lost when the subsystem $B$ is ignored; see
also Fig. \ref{fig:MAB}. The following theorem establishes a connection
between the partial entanglement and the quantum discord.
\begin{thm}
\label{thm:2}The quantum discord of a bipartite state $\rho^{AB}$
is equal to the minimal partial distillable entanglement in a von
Neumann measurement on $A$: $\delta^{\rightarrow}\left(\rho^{AB}\right)=\underset{U}{\min}P_{E_{D}}\left(U\rho_{1}U^{\dagger}\right)$.
The minimization is done over all unitaries $U$ which realize a von
Neumann measurement on $A$, and $\rho_{1}=\ket{0^{M}}\bra{0^{M}}\otimes\rho^{AB}$.\end{thm}
\begin{proof}
We note that for any state $\rho^{AB}$ the quantum discord can be
written as $\delta^{\rightarrow}\left(\rho^{AB}\right)=S\left(\rho^{A}\right)-S\left(\rho^{AB}\right)+\underset{\left\{ \Pi_{i}^{A}\right\} }{\min}\left\{ S\left(\sum_{i}\Pi_{i}^{A}\rho^{AB}\Pi_{i}^{A}\right)-S\left(\sum_{i}\Pi_{i}^{A}\rho^{A}\Pi_{i}^{A}\right)\right\} $
with the minimization over all von Neumann measurements on $A$. To
see this we start with the definition of the discord in (\ref{eq:discord}).
Then it is sufficient to show that for $p_{i}=\mathrm{Tr}\left[\Pi_{i}^{A}\rho^{AB}\Pi_{i}^{A}\right]$
and $\rho_{i}=\Pi_{i}^{A}\rho^{AB}\Pi_{i}^{A}/p_{i}$ holds $\sum_{i}p_{i}S\left(\rho_{i}\right)=S\left(\sum_{i}\Pi_{i}^{A}\rho^{AB}\Pi_{i}^{A}\right)-S\left(\sum_{i}\Pi_{i}^{A}\rho^{A}\Pi_{i}^{A}\right)$,
which can be seen by inspection using the fact that $\left\{ p_{i}\right\} $
are eigenvalues of $\sum_{i}\Pi_{i}^{A}\rho^{A}\Pi_{i}^{A}$. Using
the same arguments as in the proof of Theorem \ref{thm:1} the desired
result follows.
\end{proof}
Using Theorem \ref{thm:2} we will show that the properties (\ref{eq:deficitLO})
and (\ref{eq:deficitLO-1}) are also satisfied by the quantum discord.
Inequality (\ref{eq:deficitLO}) can be seen to be true by noting
that $E_{D}$ does not increase under LOCC and that $\Lambda_{B}$
does not change the state $\mathrm{Tr}_{B}\left[U\rho_{1}U^{\dagger}\right]$.
To see that (\ref{eq:deficitLO-1}) also holds for the quantum discord
note that, using the same arguments as in the proof of Theorem \ref{thm:1},
we can replace the distillable entanglement $E_{D}$ by the relative
entropy of entanglement $E_{R}$ in Theorem \ref{thm:2} without changing
the statement. Because of convexity of $E_{R}$ \cite{Vedral1998},
the entanglement $E_{R}^{M|A}$ is nondecreasing on average under
quantum operations on $B$: $\sum_{i}q_{i}E_{R}^{M|A}\left(\sigma_{i}^{MA}\right)\geq E_{R}^{M|A}\left(\rho^{MA}\right)$.
This implies that the partial entanglement $P_{E_{R}}\left(\rho\right)=E_{R}^{M|AB}\left(\rho\right)-E_{R}^{M|A}\left(\rho^{MA}\right)$
is nonincreasing on average under quantum operations on $B$. Using
this result we see that (\ref{eq:deficitLO-1}) also holds for the
quantum discord. 

Theorem \ref{thm:2} allows us to generalize the quantum discord to
arbitrary measures of entanglement $E$ in the same way as it was
done for the one-way information deficit in (\ref{eq:generalizedDeficit}):
\begin{equation}
\delta_{E}^{\rightarrow}\left(\rho^{AB}\right)=\underset{U}{\min}P_{E}\left(U\rho_{1}U^{\dagger}\right).\label{eq:generalizedDiscord}\end{equation}
Using the same arguments as above Eq. (\ref{eq:generalizedDiscord})
we see that the generalized quantum discord $\delta_{E}^{\rightarrow}$
satisfies the properties (\ref{eq:deficitLO}) and (\ref{eq:deficitLO-1})
for all measures of entanglement $E$ which are convex and obey (\ref{eq:LOCC}). 

So far we have only considered von Neumann measurements. In the following
we will show that our approach is also valid with an alternative definition
of the quantum discord \cite{Fanchini,Cavalcanti,Madhok}: $\delta_{\mathrm{POVM}}^{\rightarrow}\left(\rho^{AB}\right)=S\left(\rho^{A}\right)-S\left(\rho^{AB}\right)+\underset{\left\{ M_{i}^{A}\right\} }{\min}\sum_{i}p_{i}S\left(\rho_{i}^{B}\right)$,
with $\left\{ M_{i}^{A}\right\} $ being a positive operator-valued
measure (POVM) on $A$, $p_{i}=\mathrm{Tr}\left[M_{i}^{A}\rho^{AB}\right]$
and $\rho_{i}^{B}=\mathrm{Tr}_{A}\left[M_{i}^{A}\rho^{AB}\right]/p_{i}$.
The minimization over POVMs can be replaced by a minimization over
orthogonal projectors of rank one $\left\{ \Pi_{i}^{A'}\right\} $
on an extended Hilbert space $\mathcal{H}_{A'}$ with $\dim\mathcal{H}_{A'}\geq\dim\mathcal{H}_{A}$
\cite{Peres1995}. With this observation we see that all results presented
for the quantum discord also hold for the alternative definition of
the quantum discord.

In the following we will generalize our approach to multipartite von
Neumann measurements on $A$. We split the system $A$ into $n$ subsystems:
$A=\cup_{i=1}^{n}A_{i}$. A von Neumann measurement $\Lambda$ will
be called \emph{$n$-partite}, if it can be expressed as a sequence
of von Neumann measurements $\Lambda_{i}$ on each subsystem $A_{i}$:
$\Lambda\left(\rho\right)=\Lambda_{1}\left(\ldots\Lambda_{n}\left(\rho\right)\right)$.
Now we can introduce the $n$-partite one-way information deficit
$\Delta_{n}^{\rightarrow}$ and the $n$-partite quantum discord $\delta_{n}^{\rightarrow}$
as follows: \begin{eqnarray}
\Delta_{n}^{\rightarrow}\left(\rho^{AB}\right) & = & \min_{\Lambda}S\left(\Lambda\left(\rho^{AB}\right)\right)-S\left(\rho^{AB}\right),\\
\delta_{n}^{\rightarrow}\left(\rho^{AB}\right) & = & \min_{\Lambda}\left\{ S\left(\Lambda\left(\rho^{AB}\right)\right)-S\left(\Lambda\left(\rho^{A}\right)\right)\right\} \nonumber \\
 &  & -S\left(\rho^{AB}\right)+S\left(\rho^{A}\right).\end{eqnarray}
Using the same arguments as in the proof of Theorems \ref{thm:1}
and \ref{thm:2}, we see that $\Delta_{n}^{\rightarrow}$ quantifies
the minimal distillable entanglement between $M$ and $AB$ created
in an $n$-partite von Neumann measurement on $A$. $\delta_{n}^{\rightarrow}$
can be interpreted as the corresponding minimal partial distillable
entanglement $P_{E_{D}}$. We also note that this generalization includes
$n$-partite von Neumann measurements on the total system. This can
be achieved by defining $A$ to be the total system. Since $\delta_{n}^{\rightarrow}=0$
in this case, the only nontrivial quantity is the generalized information
deficit $\Delta_{n}^{\rightarrow}$. A different approach to extend
the quantum discord to multipartite settings was introduced in \cite{Modi2010}.

In this work we showed that the one-way information deficit is equal
to the minimal distillable entanglement between the measurement apparatus
$M$ and the system $AB$ which has to be created in a von Neumann
measurement on $A$. The quantum discord is equal to the corresponding
minimal partial distillable entanglement. Our approach can also be
applied to any other measure of entanglement, thus defining a class
of quantum correlation measures. This correspondence allows us to
translate certain properties of entanglement measures to corresponding
properties of quantum correlation measures. It may lead to a better
understanding of the quantum discord and related measures of quantum
correlations, since it allows us to use the great variety of powerful
tools developed for quantum entanglement. We found a class of quantum
operations which do not increase the generalized versions of the one-way
information deficit and the quantum discord. We also generalized our
results to multipartite settings. 
\begin{acknowledgments}
We thank Sevag Gharibian for interesting discussions and an anonymous
referee for constructive suggestions. We acknowledge partial financial
support by Deutsche Forschungsgemeinschaft.
\end{acknowledgments}

\paragraph*{Note added.}

Recently an alternative approach to connect the entanglement to quantum
correlation measures was presented in \cite{Piani2011}. There the
authors show that nonclassical correlations in a multipartite state
can be used to create entanglement in an activation protocol.

\bibliographystyle{apsrev4-1}
\bibliography{literature}

\end{document}